\documentclass{article}
\pdfoutput=1
\usepackage[a4paper, top=2cm,bottom=2.cm,left=2.cm,right=2.cm]{geometry}
\usepackage[english]{babel}
\usepackage[T1]{fontenc}
\usepackage{lmodern}
\usepackage[utf8]{inputenc}
\usepackage[scaled=0.95]{helvet}
\usepackage{mathptmx}
\usepackage[dvipsnames]{xcolor}
\usepackage[colorlinks, breaklinks]{hyperref}
\hypersetup{linkcolor = TealBlue, citecolor = black, urlcolor = TealBlue} 
\usepackage[font={small},labelfont={sf,bf}, margin=0cm, indention = 0cm]{caption}
\usepackage{subcaption}
\usepackage{titling}
\usepackage{authblk}
\usepackage{fancyhdr}
\usepackage{setspace}
\usepackage{sectsty}
\usepackage{microtype}
\usepackage{tikz}
\usepackage{pgf}
\usetikzlibrary{math}
\usepackage{balance}
\usepackage[nolists,nomarkers]{endfloat}
\usepackage{multicol}
\usepackage[final,defaultcolor=red]{changes}
\usepackage{csquotes}
\usepackage[style=ieee,isbn=false,url=false]{biblatex}
\newcommand\figref{Figure~\ref}

\pretitle{\centering\sffamily\LARGE\bfseries}
\posttitle{\par}

\predate{}
\postdate{}
\date{}

\allsectionsfont{\sffamily\bfseries\normalsize}

\title{PARALLELPROJ - An open-source framework for fast calculation of projections in tomography}
\author[1]{Georg Schramm}
\author[2,3]{Kris Thielemans}
\affil[1]{Department of Imaging and Pathology, Division of Nuclear Medicine, 
KU Leuven, Leuven, Belgium}
\affil[2]{Institute of Nuclear Medicine, University College London, London, UK}
\affil[3]{Centre for Medical Image Computing, University College London, London, UK}

\begin{document}

\maketitle

\begin{quote}
\textbf{\textsf{Abstract:}}

In this article, we introduce \texttt{parallelproj}, a novel open-source framework designed for efficient parallel computation of projections in tomography leveraging either multiple CPU cores or GPUs. 
This framework efficiently implements forward and back projection functions for both sinogram and listmode data, utilizing Joseph's method, which is further extended to encompass time-of-flight (TOF) PET projections.
Our evaluation involves a series of tests focusing on PET image reconstruction using data sourced from a state-of-the-art clinical PET/CT system. We thoroughly benchmark the performance of the projectors in non-TOF and TOF, sinogram, and listmode employing multi CPU-cores, hybrid CPU/GPU, and exclusive GPU mode.
Moreover, we also investigate the timing of non-TOF sinogram projections calculated in STIR
(Software for Tomographic Image Reconstruction) which recently integrated \texttt{parallelproj} as one of its projection backends.
Our results indicate that the exclusive GPU mode provides acceleration factors between 25 and 68 relative to the multi-CPU-core mode. Furthermore, we demonstrate that OSEM listmode reconstruction of state-of-the-art real-world PET data sets is achievable within a few seconds using a single consumer GPU. 
\end{quote}

\begin{multicols}{2}
  \section{Introduction}

For tomographic imaging techniques used in medicine such as X-ray computed tomography (CT), 
positron emission tomography (PET) and single photon emission tomography (SPECT), image reconstruction results 
are usually  expected within seconds or minutes after data acquisition, 
creating a severe computational challenge when reconstructing data from state-of-the-art systems 
using iterative algorithms \added{\cite{eklund_medical_2013}}.
With new scanner generations, this challenge is steadily growing, since 
(i) the data size is increasing due to higher resolution detectors and scanners with bigger field of 
view \cite{Zhang_2017}, and (ii) more advanced (iterative) reconstruction algorithms are being used that try to 
exploit more information from the acquired data, 
which usually necessitates the calculation of a huge amount of projections.
An example of the latter is the data-driven motion correction in PET \cite{Lamare_2022} where 
instead of reconstructing a single ``static frame'', many very short time frames are reconstructed and
subsequently used for motion estimation and correction.
Another example for (ii) is the combination of deep learning and tomographic image reconstruction 
\cite{Wang2018a,wang_deep_2020,Reader2021}, using, e.g. unrolled networks, 
where during training also a tremendous number of projections have to be calculated to evaluate
the gradient of the data fidelity term across a mini batch in every training epoch.

For most tomographic image reconstruction algorithms,  the bottleneck in terms of computation time is
the evaluation of a linear forward model that describes the physics of the data acquisition process.
In CT, and PET, the forward model includes the computation of many (weighted) line or volume 
integrals through an image volume, commonly called ``projections'' - which can be slow when executed 
on a single processor. 
Fortunately, for most reconstruction algorithms, the computation of projections can be executed 
in parallel on multiple processors, e.g. using multiple CPU-cores or one or more graphics processing units (GPUs). 
Note that the parallel evaluation of the adjoint of the forward model - commonly called ``back projection'' - 
is more demanding, since race conditions, where multiple threads/processes need to update the same memory location, can occur.
In recent decades, the use of GPUs for faster calculation of projections in 
tomographic imaging has been studied extensively; see, e.g. \cite{pratx_fully_2006, pratx_fast_2009, barker_gpu-accelerated_2009, cui_fully_2011, herraiz_fully_2011, herraiz_gpu-based_2011, kim_fully_2011, zhou_fast_2011, chou_accelerating_2012, kinouchi_gpu-based_2012,cui_distributed_2013, ha_gpu-accelerated_2013, markiewicz_high_2014, zhou_efficient_2014, nassiri_fast_2015, zeng_gpu-accelerated_2020} or the reviews \cite{eklund_medical_2013, depres_review_2017} for the use of GPUs in PET reconstruction.
All of these articles conclude that the time needed to calculate forward and back projections on 
state-of-the-art GPUs is usually much shorter compared to using multiple CPU-cores.

Motivated by these findings and the recent availability of very powerful low- and high-level GPU programming
frameworks such as \texttt{CUDA} and \texttt{cupy} \cite{cupy_2017}, 
we developed a new open source research framework, called \texttt{parallelproj}, for fast calculations
of \added{forward and back} projections in tomographic image reconstruction.

The objectives of the \texttt{parallelproj} framework are as follows:
\begin{itemize}
    \item To provide an open source framework for fast parallel calculation of time-of-flight (TOF) as well as non-TOF projections suited for tomographic image reconstruction in sinogram as well as listmode using multiple CPU-cores or GPUs.
    \item To provide an accessible framework that can be easily installed without the need for 
          compilation of source code on all major operating systems (Linux, Windows, and macOS).
    \item To provide a framework that can be efficiently used in conjunction with 
          \texttt{pytorch} \cite{pytorch_2019} 
          GPU arrays to facilitate research on tomographic imaging methods, including deep learning.
\end{itemize}
In light of the absence of an open-source framework that fully meets these criteria at the time of writing, this article introduces the new \texttt{parallelproj} framework
and is structured as follows: 
We first review Joseph's method for calculating projections, followed by a short overview of
the design choices and implementation of \texttt{parallelproj}.
Subsequently, we report the results of a few benchmark tests related to image reconstruction
in PET with and without TOF information using sinograms or listmode (LM) 
before ending the article with a detailed discussion and conclusion.
In this article, we focus on the performance of \texttt{parallelproj} projectors for non-TOF and TOF
PET reconstruction problems.
Note, however, that the non-TOF Joseph projectors could also be used in iterative CT reconstruction.

\begin{figure*}
\centering
\begin{tikzpicture}
\tikzmath{\x0 = 2; \y0 =2; \w = 5;} 
\draw[step=1.0,black,thick] (\x0,\y0) grid (\x0 + \w,\y0 + \w);
\draw[step=1.0,black,dotted,thin,xshift=0.5cm,yshift=0.5cm] (\x0 - 0.75,\y0 - 0.75) grid (\x0 + \w - 0.25, \y0 + \w - 0.25);
\fill[gray!5] (\x0 - 0.5, \y0 - 0.5) rectangle (\x0 -0.5 + \w,\y0 -0.5 + \w);

\tikzmath{\x0 = 1; \y0 =1; \w = 5;} 
\draw[step=1.0,black,thick] (\x0,\y0) grid (\x0 + \w,\y0 + \w);
\draw[step=1.0,black,dotted,thin,xshift=0.5cm,yshift=0.5cm] (\x0 - 0.75,\y0 - 0.75) grid (\x0 + \w - 0.25, \y0 + \w - 0.25);
\fill[gray!5] (\x0 - 0.5, \y0 - 0.5) rectangle (\x0 -0.5 + \w,\y0 -0.5 + \w);
\tikzmath{\x0 = 0; \y0 =0; \w = 5;} 
\fill[white] (\x0,\y0) rectangle (\x0 + \w,\y0 + \w);
\draw[step=1.0,black,thick] (\x0,\y0) grid (\x0 + \w,\y0 + \w);
\draw[step=1.0,black,dotted,thin,xshift=0.5cm,yshift=0.5cm] (\x0 - 0.75,\y0 - 0.75) grid (\x0 + \w - 0.25, \y0 + \w - 0.25);
\node[anchor=west] at (\x0+\w,\y0) (plane1) {\footnotesize image plane $k$};
\node[anchor=west] at (\x0+\w + 1.0,\y0 + 1) (plane2) {\footnotesize image plane $k+1$};
\node[anchor=west] at (\x0+\w + 2.0,\y0 + 2) (plane3) {\footnotesize image plane $k+2$};
\tikzmath{\xc = 1.5; \yc =2.5; \xp=1.7; \yp=2.8;}
\filldraw[gray] (\xc,\yc) circle (2pt);
\node[below] at (\xc, \yc) {\tiny$(i,j)$};
\filldraw[gray] (\xc,\yc+1) circle (2pt);
\node[above] at (\xc, \yc+1) {\tiny$(i+1,j)$};
\filldraw[gray] (\xc+1,\yc) circle (2pt);
\node[below] at (\xc+1, \yc) {\tiny$(i,j+1)$};
\filldraw[gray] (\xc+1,\yc+1) circle (2pt);
\node[above] at (\xc+1, \yc+1) {\tiny$(i+1,j+1)$};
\draw[black,thin] (\xp,\yc) -- (\xp,\yc+1);
\draw[black,thin] (\xc,\yp) -- (\xc+1,\yp);
\tikzmath{\ux=1; \uy = 0.4;}
\draw[magenta, very thick] (\xp-4*\ux,\yp-4*\uy) -- (\xp-3*\ux,\yp-3*\uy);
\draw[magenta, very thick] (\xp-4*\ux,\yp-4*\uy) -- (\xp-4*\ux,\yp-4*\uy + 0.1);
\draw[MidnightBlue, very thick] (\xp-3*\ux,\yp-3*\uy) -- (\xp-2*\ux,\yp-2*\uy);
\draw[magenta, very thick] (\xp-3*\ux,\yp-3*\uy) -- (\xp-3*\ux,\yp-3*\uy + 0.1);
\draw[magenta, very thick] (\xp-2*\ux,\yp-2*\uy) -- (\xp-1*\ux,\yp-1*\uy);
\draw[magenta, very thick] (\xp-2*\ux,\yp-2*\uy) -- (\xp-2*\ux,\yp-2*\uy + 0.1);
\draw[MidnightBlue, very thick] (\xp-1*\ux,\yp-1*\uy) -- (\xp-0*\ux,\yp-0*\uy);
\draw[magenta, very thick] (\xp-1*\ux,\yp-1*\uy) -- (\xp-1*\ux,\yp-1*\uy + 0.1);
\draw[MidnightBlue, very thick] (\xp+5.5*\ux,\yp+5.5*\uy) -- (\xp+6*\ux,\yp+6*\uy);
\draw[magenta, very thick] (\xp+6*\ux,\yp+6*\uy) -- (\xp+7*\ux,\yp+7*\uy);
\draw[magenta, very thick] (\xp+6*\ux,\yp+6*\uy) -- (\xp+6*\ux,\yp+6*\uy + 0.1);
\draw[magenta, very thick] (\xp+7*\ux,\yp+7*\uy) -- (\xp+7*\ux,\yp+7*\uy + 0.1);
\node at (\xp-4*\ux,\yp-4*\uy) (nodeA) {};
\node at (\xp-3*\ux,\yp-3*\uy) (nodeB) {};
\node at (\xp-2*\ux,\yp-2*\uy) (nodeC) {};
\node at (\xp+6*\ux,\yp+6*\uy) (nodeD) {};
\node at (\xp+7*\ux,\yp+7*\uy) (nodeE) {};
\draw[magenta, very thick] (nodeA) -- (nodeB) node [midway, above, sloped] (TextNode) {\tiny TOF bin $t$};
\draw[MidnightBlue, very thick] (nodeB) -- (nodeC) node [midway, above, sloped] (TextNode) {\tiny TOF bin $t+1$};
\draw[magenta, very thick] (nodeD) -- (nodeE) node [midway, above, sloped] (TextNode) {\tiny TOF bin $t+n$};
\filldraw[Peach] (\xp,\yp) circle (2pt);
\node[anchor=north,align=center,font={\footnotesize}] at (\xp-4*\ux,\yp-4*\uy) (ray) {ray with or without\\subdivision into TOF bins};
\end{tikzpicture}
\caption{Illustration of Joseph's method for projecting rays through voxel images. In a ray-driven approach
the image volume is traversed plane by plane along a principal direction. At every plane, the intersection point
between the ray and image plane is calculated (orange dot). The contribution of the four nearest voxels (gray dots)
to the line integral is modeled using bi-linear interpolation weights. 
The method can also be easily extended to compute TOF-weighted projections using a subdivision of the ray into
TOF bins and by evaluation of a TOF kernel. 
See text and \cite{Joseph1982} for more details. Figure not drawn to scale.}
\label{fig:joseph}
\end{figure*}
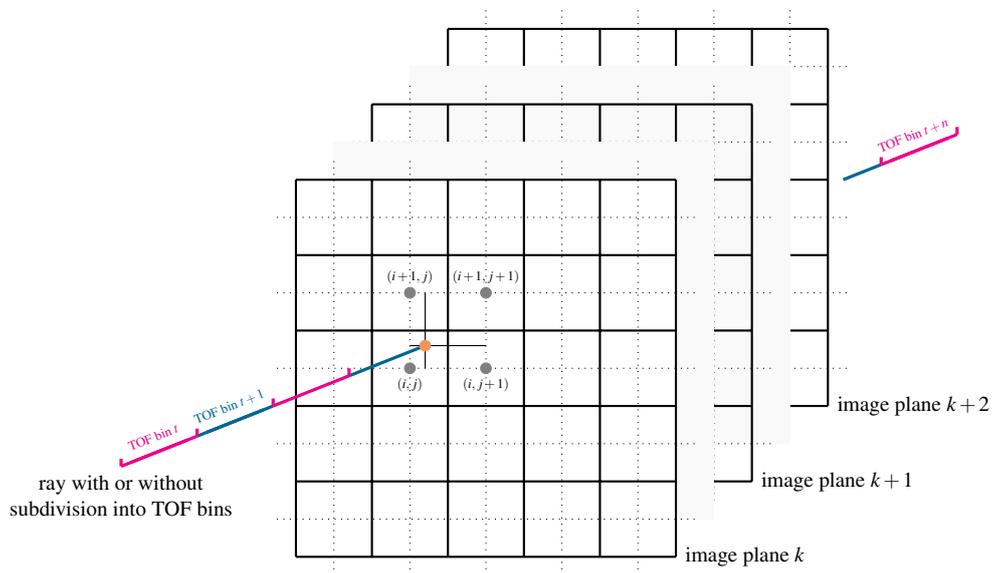

\section{Materials and Methods}

\subsection{Joseph's method for projecting rays through voxel images}

Besides Siddon's method \cite{siddon_fast_1985}, \added{Wu's method \cite{Wu1991}} and the distance-drive method \cite{DeMan2004}, Joseph's
method \cite{Joseph1982} is a very efficient and popular way to calculate projections in transmission and emission
tomography.
\added{Rahmin et al. \cite{Rahmim2005} have shown that while being only 20\% slower than Siddon's method, Joseph's method
      leads to superior image quality in listmode PET reconstructions.}
The basic idea of the original Joseph method for calculating line integrals, that can be used
to model non-TOF projections, is shown in \figref{fig:joseph}.
For a given ray, the algorithm first determines the principal direction in the image 
that is most parallel to the ray
and then steps through the image volume plane by plane along this principal direction. 
In every plane, the intersection point between the ray and the plane is calculated and the contribution 
of the image at that point to the line integral is approximated using bi-linear interpolation of the four nearest neighbors 
around the intersection point.
In other words, only the four nearest neighboring voxels are contributing to the line integral and
their contributions are given by the bi-linear interpolation weights.
Finally, the contributions of all planes are added and corrected for the incidence angle of the ray, see \cite{Joseph1982} for more details.

An extension of the original Joseph method to calculate TOF-weighted projections is straightforward.
For every voxel contributing to the line integral and every TOF bin along the ray, a TOF weight can be
computed by evaluating a TOF kernel that is a function of the Euclidean distance between the voxel
and the center of the TOF bin.
The TOF kernel can be e.g. modeled as a Gaussian kernel - representing the TOF uncertainty of the detection system - convolved with a rectangular function - representing the width of the TOF bin, resulting in the evaluation of the difference of two error functions.

\subsection{Design principles and implementation details}

The application programming interface (API) to the \texttt{parallelproj} framework was designed such that:
\begin{itemize}
    \item The input to the low level projector functions are as generic as possible.
          In practice, that means that these functions take a list of coordinates representing the
          start and end point of the rays to be projected as input, making the low-level functions
          agnostic to specific scanner geometries (or symmetries). Thus, any scanner geometry
          can be modeled.
    \item Projections can be performed in non-TOF or TOF mode.
    \item In the TOF mode, optimized projections for sinogram and listmode are available.
          In the former, the contributions to all available TOF bins along a ray are computed while
          traversing the image volume plane by plane, whereas in the latter only the contribution to
          one specific TOF bin (the TOF bin of a given listmode event) is evaluated.
    \item The back projections are the exact adjoint of the forward projections (matched forward and 
          back projections).
\end{itemize}
Parallelization across multiple processors was implemented in two different ways. 
To enable parallelization across multiple CPUs, a first version of the \texttt{parallelproj} library was implemented using C and 
OpenMP \cite{openmp_1998} (\texttt{libparalleproj\_c}). 
Furthermore, the exact same projector functions were implemented in \texttt{CUDA} to enable parallelization 
on one or multiple GPUs (\texttt{libparalleproj\_cuda}).
In the \texttt{CUDA} version, the input data is first transferred from the host to all available GPU(s)
followed by the parallel execution of the projection kernels.
After the kernel execution, the result is transferred back to the host.
To handle race conditions, all implementations use atomic add operations in the back projections.

\subsection{Availability of source code and precompiled libraries}

\texttt{parallelproj} is an open source project and its source code is available at 
\url{https://github.com/gschramm/parallelproj} under an MIT license.
In addition to the sources, we also offer precompiled libraries
(\texttt{libparallelproj\_c} and \texttt{libparallelproj\_cuda})
for all major operating systems (Linux, Windows,
and macOS) and various recent \texttt{CUDA} versions using the 
\href{https://github.com/conda-forge/parallelproj-feedstock}{\texttt{conda-forge}} 
package manager\footnote{The  \texttt{CUDA} version of the \texttt{parallelproj} 
library is not available for macOS.}.
Depending on the presence or absence of supported CUDA devices and drivers, \texttt{conda-forge} automatically
installs the matching library type.
In addition to the precompiled libraries, the \texttt{parallelproj} package also includes the source file of the \texttt{CUDA} projection kernels such that they can be directly executed on GPU arrays using frameworks that allow for just-in-time compilation of \texttt{CUDA} kernels such as, e.g. \texttt{cupy} \cite{cupy_2017}.
Moreover, \texttt{parallelproj} also includes a minimal python interface to all projection functions that is compatible with the Python array API standard, enabling efficient projections and back projections of various compatible array classes (e.g. \texttt{numpy}, \texttt{cupy}, \texttt{pytorch} tensors).

\subsection{\texttt{parallelproj} computation modes}

Using the two aforementioned projection libraries, as well as the \texttt{CUDA} projections kernels,  
projections can be performed in the following three different computation modes:
\begin{enumerate}
    \item \textbf{CPU mode:} Forward and back projections of image volumes (arrays) stored on the host (CPU) 
           can be performed using \texttt{libparallelproj\_c} where parallelization across all available 
           CPU cores is performed  using OpenMP.
    \item \textbf{hybrid CPU/GPU mode:} Forward and back projections of image volumes (arrays) stored on 
           the host can be performed using \texttt{libparallelproj\_cuda} involving data transfer from the host to all available GPUs,
           execution of projection kernels on the GPUs, and transfer of the results back to the host.
    \item \textbf{direct GPU mode:} Forward and back projections of image volumes (arrays) stored on a GPU can
           be performed by direct execution of the projection kernels using a framework that supports 
           just-in-time compilation of \texttt{CUDA} kernels, such as \texttt{cupy} \cite{cupy_2017}. 
           In contrast to the hybrid CPU/GPU mode, memory transfer between host and GPU is avoided\footnote{A requirement for the use of the direct GPU mode is the presence of enough GPU memory to store the input/output images and projections.}.
\end{enumerate}

\subsection{Integration of \texttt{parallelproj} into STIR\label{sec:STIR}}

Software for Tomographic Image Reconstruction (STIR) is open source software for PET and SPECT reconstruction \cite{thielemans2012STIRSoftwareTomographic,fuster2013IntegrationAdvanced3D}. It is a well-established tool for
research in scanner modeling and iterative reconstruction methods. Its modular design in C++ allows integrating external
components such as projectors. We integrated \texttt{parallelproj} into STIR since version 5.0 as a user-selectable 
projector such that
STIR users can benefit from the high performance of the \texttt{parallelproj}. STIR's \href{https://github.com/conda-forge/stir-feedstock}{\texttt{conda-forge} recipe}
depends on \texttt{parallelproj} and therefore installs the GPU or CPU version accordingly.
Moreover, as STIR forms the basis
for the PET and SPECT support in the open source Synergistic Image Reconstruction Framework (SIRF) \cite{ovtchinnikov2020SIRFSynergisticImage},
this was modified to allow calling \texttt{parallelproj} from SIRF as well, making \added{parallelproj} usable for \added{SIRF's} advanced algorithms,
including motion correction \cite{brown2021MotionEstimationCorrection}.

STIR uses \texttt{parallelproj} if the latter's libraries are found by \texttt{CMake} at compilation time. Currently, STIR always uses
\texttt{libparallelproj\_cuda} if \added{present}, and \texttt{libparallelproj\_c} otherwise. As \texttt{parallelproj} needs the end-points
of the lines of responses, these are computed by the STIR interface based on its normal modeling of scanner geometry, defaulting to
cylindrical scanners but since recently also accommodating block-cylindrical and arbitrary crystal locations. These computations are
performed once at set-up time, and end-points are stored in \texttt{std::vector}s suitable for passing to the low-level routines of
\texttt{parallelproj}. Since STIR 5.2, these arrays are filled in parallel using OpenMP, reducing the set-up time. However, this set-up time is not
included in the timings below.

As STIR's data-structures store sinograms and images in CPU memory, this interface uses the hybrid CPU/GPU mode of \texttt{parallelproj}.
STIR's design for projectors is optimized for low-memory requirements, projecting only small chunks of sinogram data at the time,
using OpenMP or the Message Passing Interface\footnote{\url{https://www.mpi-forum.org/docs/mpi-4.0/mpi40-report.pdf}} (MPI)  for non-shared memory architectures.
GPU computations however have best performance on larger data-sets. Therefore, the current implementation uses temporary objects
to store the result of the forward projection, and data is then copied as necessary. This is similar to the previous integration of
\texttt{NiftyPET} \cite{markiewicz_high_2014} into STIR. This creates an extra (small) overhead, which
could be avoided in the future.

At the time of writing, \added{neither} the TOF nor listmode projectors of \texttt{parallelproj}  have been integrated into STIR.
We hope to complete this in the near future.

\subsection{Benchmark tests}

To evaluate the performance of the \texttt{parallelproj} projectors using the computation modes described above,
we implemented a series of benchmark tests. 
All tests are related to a PET image reconstruction task and used the geometry and properties 
of a state-of-the-art GE Discovery MI TOF PET/CT scanner \cite{Hsu2017} with 20\,cm axial FOV. 
This scanner has 36 detector ``rings'', where each ``ring'' has a radius of 380\,mm and consists 
of 34 modules containing 16 detectors each such that there are 16$\times$34$\times$36 = 19584 detectors in total.
A non-TOF emission sinogram for this scanner without any data size reduction (``span 1") has 415 radial elements, 272 views, and 1292 planes,
meaning that for a full non-TOF sinogram projection, 415$\times$272$\times$1292 = 146e6 line 
integrals have to be evaluated.
For TOF data, each line of response (LOR) is subdivided into 29 TOF bins using a TOF bin width of 169\,ps (25.4\,mm).
The reported TOF resolution of the scanner is 385\,ps (57.7\,mm) FWHM \cite{Hsu2017}.
In the TOF projectors of \texttt{parallelproj}, the Gaussian TOF kernel is truncated 
beyond $\pm$3 standard deviations.

To evaluate the performance of \texttt{parallelproj} for projections in sinogram mode, 
we measured the time needed for a forward and back projection of a span 1 subset sinogram containing
8 equally spaced views in non-TOF and TOF mode. 
This is equivalent to the projection work required for an OSEM subset update \cite{Dempster1977, Lange1984, Hudson1994} using 34 subsets in total, 
a setting that is used in many clinical reconstructions.
Since it is known that the in-memory data order severely affects the computation time, especially
on \texttt{CUDA} devices, we varied the order of the spatial axis of the sinogram, 
as well as the order of the image axis relative to the axial direction of the scanner (symmetry axis mode).
In the sinogram order mode ``PVR'', the radial direction increased the fastest and the
plane direction increased the slowest in memory.
For the sinogram order mode ``VRP'', the plane direction increased the fastest and the view direction the slowest.
By varying the order of the image axis, 
we could test the impact of different image volume memory layouts. 
For example, the symmetry axis mode ``2'' meant that the image volume memory increased the fastest
in the axial direction of the scanner, while ``0'' or ``1'' meant that one of the transaxial
directions increased the fastest.
To test the integration of \texttt{parallelproj} projection libraries into STIR, 
the same non-TOF projection benchmarks tests were repeated using the timing tool included in STIR
version 5.2.
These tests were only performed in the hybrid CPU/GPU mode, since the exclusive GPU mode is currently not available in STIR.
Note that the current STIR integration uses the ``PVR'' and ``2'' symmetry axis mode.
\added{Finally, we also compared the performance of the \texttt{parallelproj} projectors in pure GPU and hybrid CPU/GPU mode with
the performance of the GPU projectors included in the \texttt{NiftyPET} python package v2.0.0 \cite{Markiewicz2018} using a complete forward
and back projections of non-TOF sinograms of the Siemens mMR \cite{Delso2011}.
Since NiftyPET uses a span 11 sinogram, and parallelproj so far only supports span 1 sinograms,
we artificially limited the maximal ring difference in this parallelproj test to 7
to obtain a sinogram with approximately the same number of planes
(NiftyPET sinogram 837 planes, parallelproj sinogram 904 planes).
In all cases, an image of shape (344, 344, 127) with a voxel size of (2.08\,mm, 2.08\,mm, 2.03\,mm) was used.}

In addition to the sinogram projection tests, we also evaluated the performance of \texttt{parallelproj}
for non-TOF and TOF projections in listmode as a function of the number of acquired listmode events.
Instead of randomly generating the event coordinates, listmode events from an acquisition of a
NEMA image quality phantom were used, which guaranteed a more realistic event distribution.  
In contrast to the projections in sinogram mode, where the ray directions and memory access
are somehow ordered, they are random for unsorted listmode data.
Similarly to the sinogram tests, the symmetry axis of the scanner was also varied.
For all sinogram and listmode projection benchmarks, the coordinates of all LOR start and 
endpoints were precalculated such that the overhead of calculating the LOR coordinates was not
included in these tests.
All benchmarks were repeated 10 times and the mean and standard deviation of the results
were calculated and visualized.

Finally, we also measured the time needed for a complete listmode OSEM iteration using
34 subsets as a function of the number of listmode events in the NEMA acquisition.
The raw listmode data, including 40 million prompt events, as well as all quantitative corrections needed for reconstruction of the NEMA phantom acquisition are available online at \url{https://doi.org/10.5281/zenodo.8404015}.

All tests used an image of size (215,215,71), an isotropic voxel size of 2.78\,mm, and  
were performed on a workstation including an AMD Ryzen Threadripper PRO 3955WX 16 core
32 thread CPU with 256\,GB RAM, and an NVIDIA GeForce RTX 3090 GPU with 24\,GB RAM
on Ubuntu 22.04 LTS using \texttt{CUDA} v11.2 and \texttt{parallelproj} v1.5.0.
\added{Note that \texttt{parallelproj}'s projectors support also non-isotropic voxel sizes.}
The scanner geometry, as well as the list mode OSEM algorithm, was implemented in a 
minimal proof-of-concept Python package available at \url{https://github.com/gschramm/parallelproj-benchmarks},
except for the STIR benchmark, where STIR's normal geometric modelling was used.
For the CPU and hybrid CPU/GPU mode, Python's \texttt{ctypes} module is used to project \texttt{numpy}
arrays stored in CPU (host) memory using a minimal interface to the low level projection functions
defined in \texttt{libparallelproj\_c} and \texttt{libparallelproj\_cuda}.
In GPU mode, the \texttt{CUDA} projection kernels were just in time compiled and directly 
executed on \texttt{cupy} GPU arrays.
Due to the interoperability between \texttt{numpy} and \texttt{cupy} the same high-level 
listmode OSEM implementation could be used for both modes.
Note that in the latter, all operations needed for the OSEM update were executed directly on the \texttt{cupy} 
GPU arrays, eliminating any memory transfer between the host and the GPU during OSEM updates.
In all listmode OSEM reconstructions, a shift-invariant image-based resolution model was used, 
including a 3D isotropic Gaussian kernel of 4.5\,mm FWHM.


\section{Results}

\begin{figure*}
      \centering
      \includegraphics[width=\textwidth]{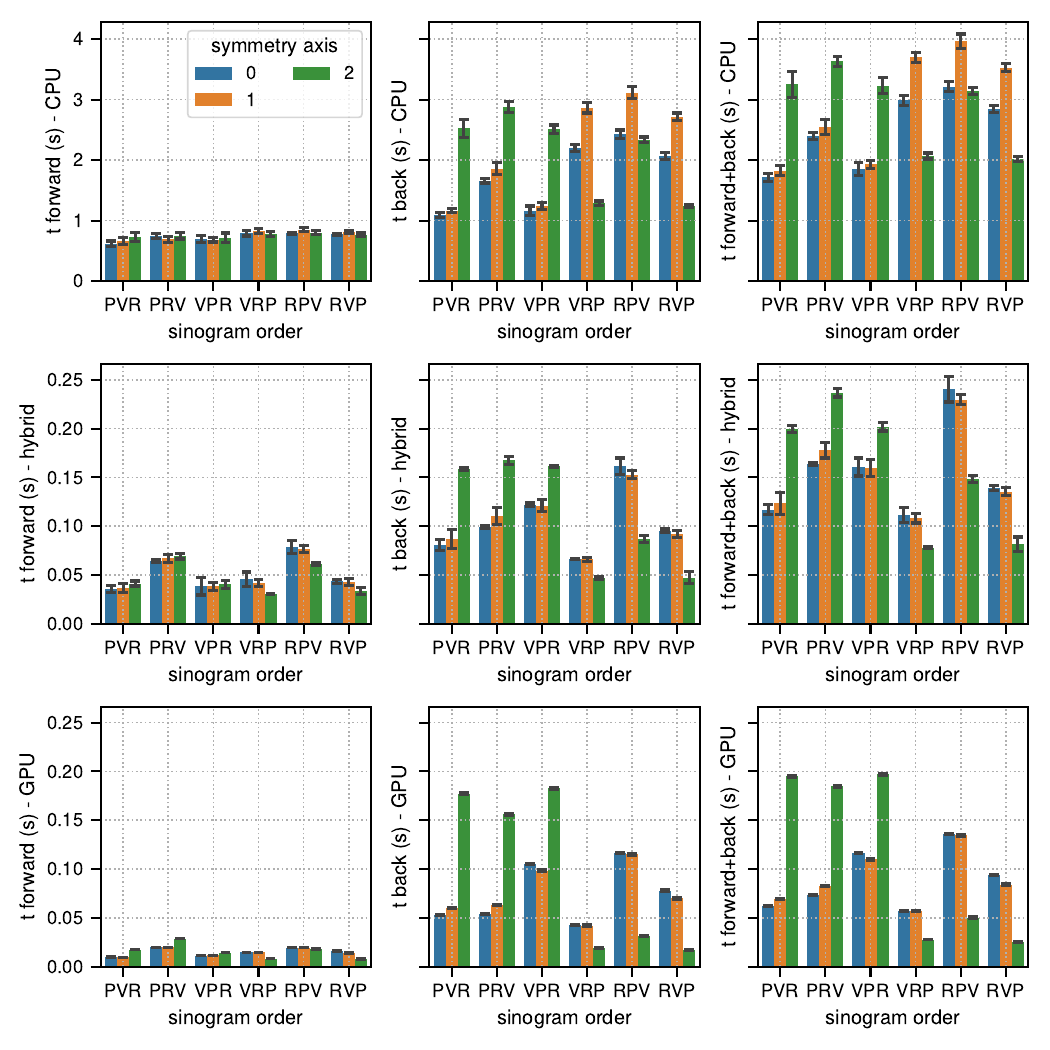}
      \caption{Results of the non-TOF sinogram benchmark tests. The non-TOF subset sinogram
            contained 415 radial elements, 8 views and 1292 planes (1 out of 34 subsets).
            The image used in these tests contained (215,215,71) voxels with an isotropic voxel size of 2.78\,mm.
            The mean and the standard deviation estimated from 10 runs are represented by the colored bars
            and the black error bars, respectively. Note the different limits on the y axes.
            The top, middle, and bottom row show the results for (multi-core) CPU, hybrid CPU/GPU and pure GPU mode,
            respectively. The left, middle, and right columns show the timing results for forward, back and combined forward and back projections, respectively.
            \added{For comparison, the time needed to calculate the same forward and back projection using the \texttt{parallelproj} projectors
                  in hybrid CPU/GPU mode integrated into STIR was 0.051\,s and 0.169\,s, respectively (see text).}}
      \label{fig:nontofsino}
\end{figure*}

\begin{figure*}
      \centering
      \includegraphics[width=\textwidth]{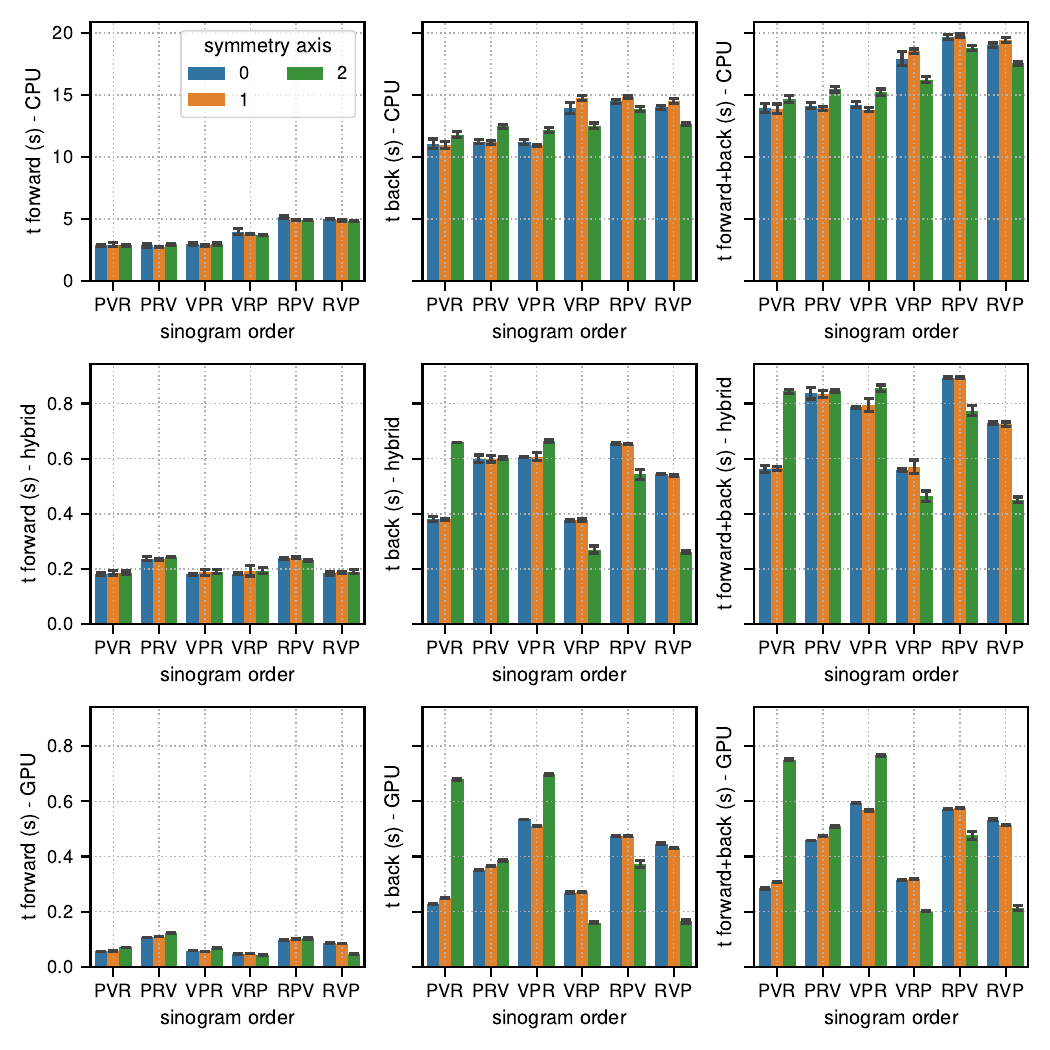}
      \caption{Same as Fig.~\ref{fig:nontofsino} for the results of the TOF sinogram benchmark tests. The TOF subset sinogram
            contained 415 radial elements, 8 views, 1292 planes and 29 TOF bins with a width of 169\,ps.
            The modeled TOF resolution was 375\,ps}
      \label{fig:tofsino}
\end{figure*}

\begin{figure*}
\centering
\includegraphics[width=\textwidth]{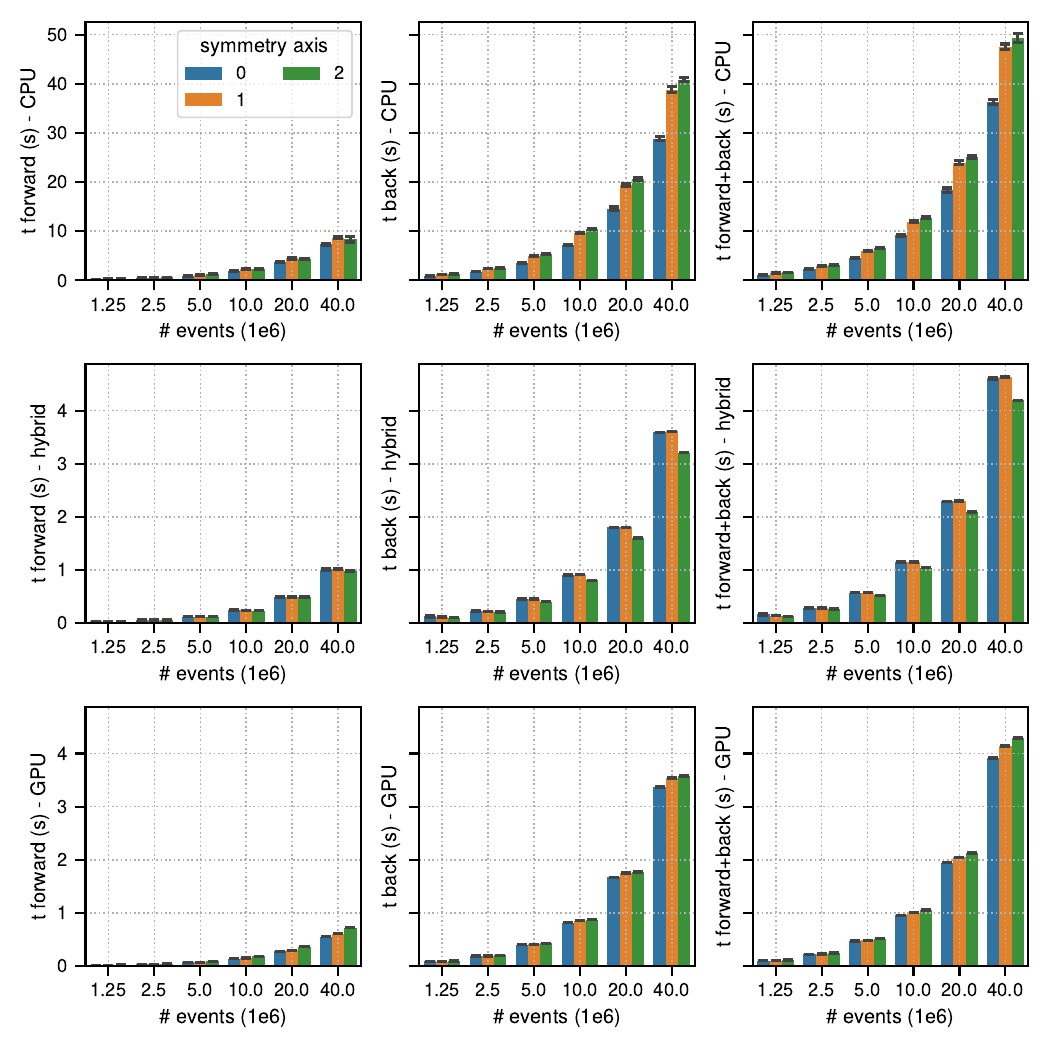}
\caption{Same as Fig.~\ref{fig:nontofsino} for results of the non-TOF listmode benchmark tests for different number of listmode events.
Note the different limits on the y axes and that the x-axis scale is non-linear.}
\label{fig:nontofLM}
\end{figure*}

\begin{figure*}
\centering
\includegraphics[width=\textwidth]{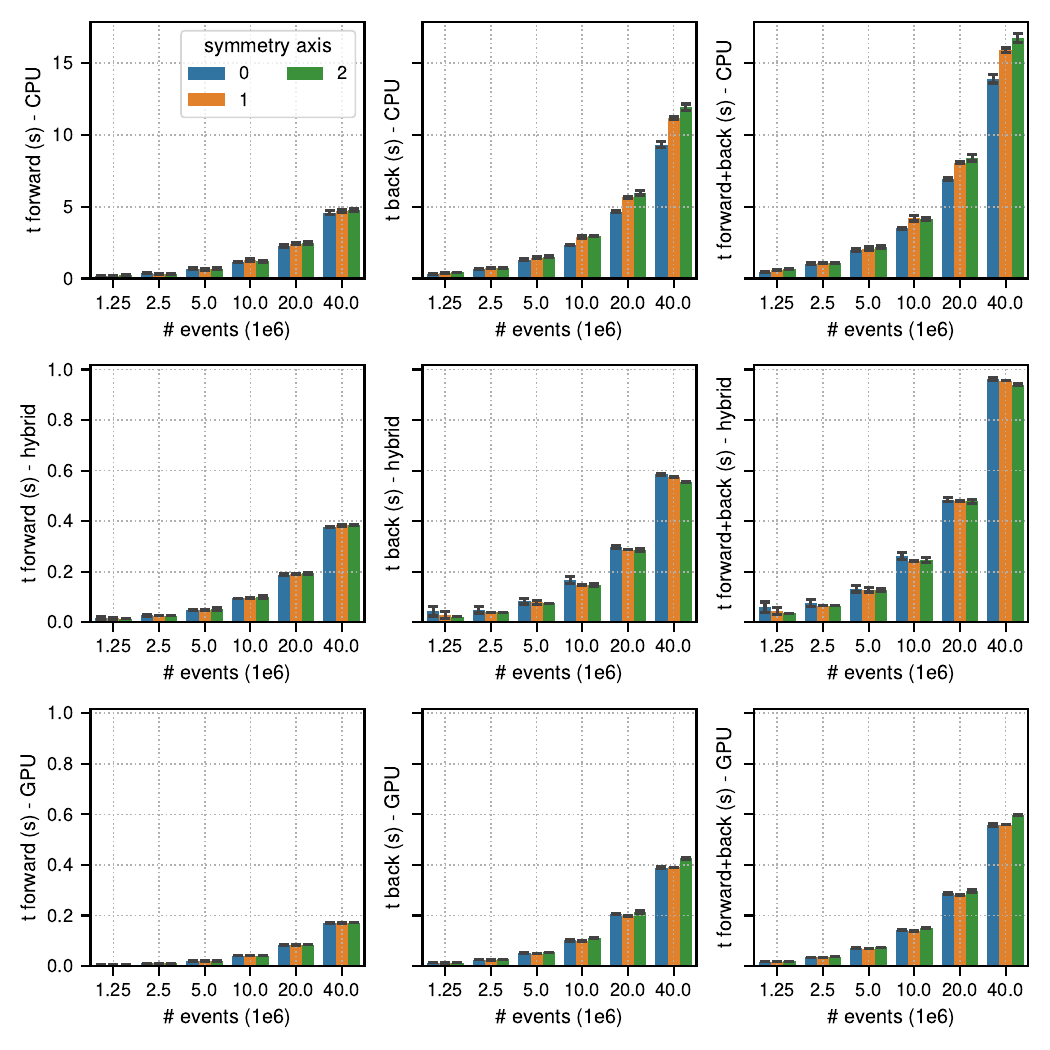}
\caption{Same as Fig.~\ref{fig:nontofsino} for results of the TOF listmode benchmark tests for different number of listmode events.
Note the different limits on the y axes and that the x-axis scale is non-linear.}
\label{fig:tofLM}
\end{figure*}

\begin{figure*}
\centering
\includegraphics[width=\textwidth]{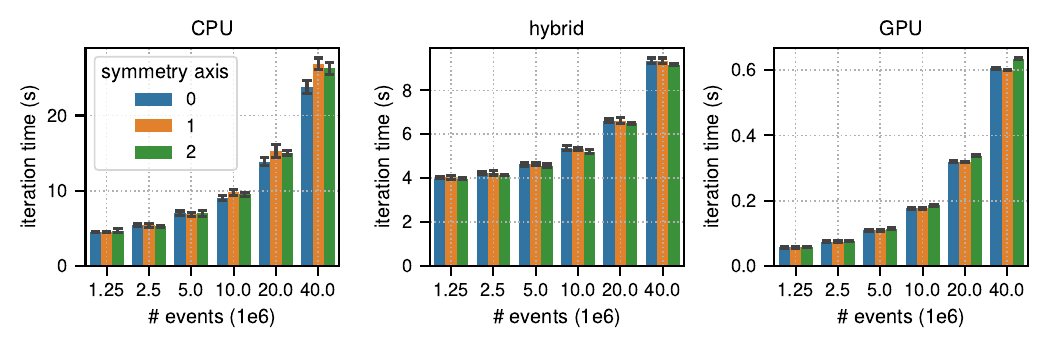}
\includegraphics[width=0.7\textwidth]{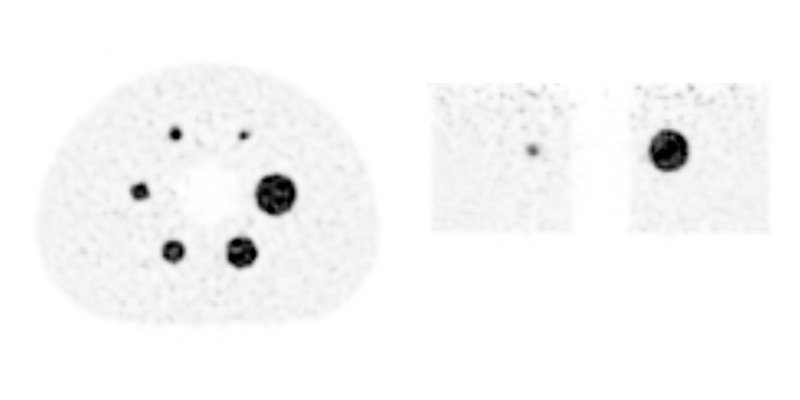}
\caption{(top) Results for the timing of a complete LM OSEM iteration including 34 subset updates
for the NEMA image quality phantom acquisition.
The image used in these tests contained (215,215,71) voxels with an isotropic voxel size of 2.78\,mm. 
The mean and the standard deviation estimated from 6 iterations are represented by the colored bars
and the black error bars, respectively. Note the different limits on the y axes and that the
x-axis scale is non-linear.
(bottom) Transaxial and coronal slice of a listmode OSEM reconstruction of the NEMA image 
quality phantom with 40e6 events after 6 iterations with 34 subsets using a 
standard Gaussian post filter of 4\,mm FWHM.
Note that for better visibility, the reconstructed image was cropped to the center portion of 
the transaxial FOV.}
\label{fig:nema_lm_osem}
\end{figure*}

Figures~\ref{fig:nontofsino} and \ref{fig:tofsino} show the results of the sinogram benchmarks in
non-TOF and TOF mode, respectively.
In non-TOF mode, the best results in terms of the summed time needed for the forward and back projection of
one subset sinogram were
(compute mode, sinogram order mode, scanner symmetry axis): 1.71\,s for (CPU, PVR, 0), 0.078\,s for (hybrid CPU / GPU, VRP, 2) and 0.025\,s for (GPU, RVP, 2), meaning that the pure GPU mode was
approximately 68x faster than the CPU mode and 3.1x faster than the hybrid mode.
Benchmarking the timing of the same projections using \texttt{paralleproj} integrated into STIR using the hybrid CPU/GPU mode, revealed very similar performance - 0.051\,s for forward projection and 0.169\,s for back projection - as compared to the corresponding results shown in the middle row of Figure~\ref{fig:nontofsino}.

\begin{table}
      \begin{center}
            \begin{tabular}{l c c}
                  \hline
                  projector                        & t forward (s) & t back (s) \\ \hline
                  parallelproj GPU mode            & 0.21          & 0.43       \\
                  parallelproj hybrid CPU/GPU mode & 0.62          & 0.95       \\
                  NiftyPET                         & 0.68          & 1.56       \\ \hline
            \end{tabular}
      \end{center}
      \caption{\added{Timing results for the forward and back projection of a non-TOF sinogram of the Siemens mMR using
                  parallelproj's GPU and hybrid CPU/GPU mode, as well as NiftyPET's projectors. See text for details.}}
      \label{tab:nifty}
\end{table}

\added{Table \ref{tab:nifty} shows the timing results for the forward and back projection of a non-TOF sinogram of the Siemens mMR and demonstrates that the timing performance of \texttt{parallelproj} in both GPU compute modes is slightly superior to the NiftyPET's GPU projector.}

In TOF mode, the corresponding results were: 13.88\,s for (CPU, PVR, 1), 0.45\,s for 
(hybrid CPU/GPU, RVP, 2), and 0.21\,s for (GPU, RVP, 2), which means that the pure GPU mode was
approximately 66x faster than the CPU mode and 2.1x faster than the hybrid mode.
As expected, especially for the back projections where atomic operations are used, the memory
order in the sinogram as well as in the image has a substantial impact on the results.
The ratios between the fastest and slowest results for the combined projection times in terms of sinogram order and symmetry axis (non-TOF, TOF mode) were: 
(2.3, 1.4) in the CPU mode, (3.1, 2.0) in the hybrid mode, and (7.6, 3.6) in the GPU mode.

Figures~\ref{fig:nontofLM} and \ref{fig:tofLM} show the results of the listmode benchmarks in
non-TOF and TOF mode, respectively.
For 40e6 events, the best results in terms of time needed for the forward and back projection (non-TOF, TOF) were: (36.3, s, 13.9, s) in CPU mode, (4.19\,s, 0.94\,s) in hybrid mode and (3.91\,s, 0.56\,s) in GPU mode.
For 1.25e6 events, the best results in terms of time needed for the forward and back projection
(non-TOF, TOF) were: (1.13\,s, 0.45\,s) in CPU mode, (0.13\,s, 0.035\,s) in hybrid mode, and
(0.1\,s, 0.017\,s) in GPU mode.
In non-TOF mode, the pure GPU mode was approximately 9.3x faster than the CPU mode and 1.3x faster than 
the hybrid mode.
In TOF mode, the pure GPU mode was approximately 24.8x faster than the CPU mode and 1.7x faster than 
the hybrid mode.
In contrast to the sinogram benchmark results, the impact of the scanner symmetry axis direction was small.
In the CPU and GPU mode, the increase in projection time as a function of the number of list-mode events was almost perfectly linear.
In hybrid mode at low number of events, the scaling was non linear due to the overhead caused by the time needed for memory transfer.

\figref{fig:nema_lm_osem} shows the results for the timing of a complete TOF listmode OSEM
iteration, including 34 subset updates, as well as a reconstruction of the NEMA image quality
phantom data set using 40e6 total prompt events.
The best results for (40e6, 1.25e6) events were: (23.82\,s, 4.47\,s)  in CPU mode, 
(9.17\,s, 3.97\,s) in hybrid mode, and (0.60\,s, 0.057\,s) in GPU mode which means that for
40e6 events the pure GPU mode was approximately 40x faster than the CPU mode and
15x faster than the hybrid mode. 

\section{Discussion}

All results shown in our article demonstrate once more that parallel computation of 
forward and back projections using a state-of-the-art GPU is substantially faster compared 
to parallelization using OpenMP on a state-of-the-art multicore CPU system.
Certainly, the achievable GPU acceleration factor strongly depends on the computational problem
itself (e.g. sinogram or listmode reconstruction) and the problem size. 
In our non-TOF and TOF sinogram and listmode benchmark tests, we observed GPU acceleration factors between 25 and 68.

One important aspect that emerged from our sinogram benchmark tests - where the 
projection data and memory access is ordered - is the fact that the projection times varied substantially when using 
different memory layouts (up to a factor of 7.6 in the GPU mode).
This can be understood by taking into account that the amount of race conditions that
are created during the back projection within a thread block heavily depends on the
order and possible intersections of rays to be projected within that block.
Note that in pure GPU mode, the time needed for sinogram forward projections
also varied substantially across the different memory layouts, which is probably due
to the way image memory is accessed and cached on \texttt{CUDA} GPUs.

Another interesting observation is the fact that in all compute modes
the time needed to calculate TOF sinogram projections was much longer than the times needed
to calculate non-TOF sinogram projections, whereas the situation was reversed in listmode.
For TOF sinogram projections, more floating point operations have to be computed compared
to non-TOF sinogram projections due to the evaluations of the TOF kernels between the
contributing voxels and a number of TOF bins.
In listmode, however, the computational work needed to project a TOF event is much
lower compared to projecting a non-TOF event. 
This is the case because a TOF listmode event detected in a specific TOF bin is only affected by a few
voxels along the complete LOR in the image, where the number of affected voxels is inversely 
proportional to the TOF resolution of the scanner.
That in turn means that with scanner TOF resolutions becoming better and better, the gap
between the TOF projection times in sinogram and listmode will become bigger and bigger,
strongly favoring listmode processing.
According to our experience, projection times in listmode are already much faster for most
standard clinical acquisitions (except for very long static brain scans with high affinity
tracers) on current PET systems with TOF resolutions between 250-400\,ps.\footnote{An alternative way
to further accelerate sinogram-based reconstructions is the use of dedicated sinogram rebinning
techniques.}
Extrapolating the timing results of one complete OSEM listmode iteration of an
acquisition with 40e6 counts in \figref{fig:nema_lm_osem}, clinical listmode OSEM 
reconstructions of a single bed position of a standard static FDG whole-body acquisition
using PET scanner with 20-25\,cm axial FOV seem to be possible in a couple 
of seconds and could even be faster than the acquisition time.\footnote{This 
is obviously only true if all other necessary corrections,
such as scatter estimation, can be performed very quickly as well.
For PET scanners with a very long axial field of view (much higher sensitivity), 
the reconstruction times could be substantially longer.}

A somewhat unexpected result was the fact that the gap in the TOF projection times between
hybrid CPU/GPU and pure GPU mode was much bigger when timing the execution of a 
complete listmode OSEM iteration compared to the pure projection benchmark test when
reconstruction 40e6 counts (approximately a factor of 15 vs a factor of 1.7, respectively).
After detailed profiling of a listmode OSEM iteration in hybrid mode, it became obvious
that the total time spent for the 34 subset listmode forward and back projections (ca. 1.2\,s) was
short compared to the time needed to calculate all other operations necessary for the OSEM update.
Profiling revealed that calculating all 68 Gaussian convolutions needed for image-based resolution modeling -
performed on the CPU in hybrid compute mode - took approximately 2.3\,s.
An interesting lesson to be learned is that once very fast GPU-based projectors are used,
it should always be double-checked whether other computational steps of any algorithm become new bottlenecks.

\added{A natural prerequisite for running sinogram OSEM reconstruction is the availability of enough GPU memory
to store the complete image volume, the emission sinogram, the forward projection and the contamination sinogram.
For TOF PET scanners with an 25\,cm axial FOV and 400\,ps TOF resolution, this means that ca. 40-50\,GB
of GPU memory is required which is available on state-of-the art server GPUs, but can be challenging for consumer GPUs.
Morever, these memory requirements increase even further for PET systems with longer axial FOV and better TOF resolution.
Note, however, that for systems with state-of-the-art TOF resolution,
the memory requirements can be severely reduced when running OSEM in listmode.
Moreover, the hybrid CPU/GPU mode of \texttt{parallelproj} allows ``chunk-wise'' calculations of projections
and supports the use of multiple GPUs to be able to reconstruct sinogram data from long axial FOV PET systems.}

An important limitation of our study is the fact that we only implemented and benchmarked Joseph's projection method.
Compared to other methods such as the distance-driven method, multiray models, or tube-of-response
models, Joseph's method is inherently faster.
Consequently, projection times are expected to be somewhat slower for more advanced projectors,
but a detailed investigation of more advanced projectors is beyond the scope of this work and 
left for future research. \footnote{Since 
\texttt{parallelproj} is an open-source project, contributions of or discussions on more advanced
projectors from the reconstruction community are more than welcome.}
Note, however, that according to our experience, combining Joseph's method with an image-based and / or 
sinogram-based resolution model can produce high-quality PET reconstructions.

Without a doubt, it is possible to further optimize the implementation of the \texttt{parallelproj}
projectors, especially the \texttt{CUDA} implementation. 
As an example, we have decided not to use \texttt{CUDA}'s texture memory, which could substantially accelerate the image 
interpolations needed in the Joseph forward projections, or be also used to interpolate TOF kernel values based on a 1D lookup 
table which would also allow the use of non-Gaussian TOF kernels \added{\cite{Efthimiou2020}}.
The main reason for not using texture memory is the fact that it would only accelerate the
forward projections since writing into texture memory is not possible and because reconstruction times 
are usually dominated by the back projections.
Another way to further improve the listmode projection times is to pre-sort the listmode events
to minimize race conditions during back projection, as e.g. shown in \cite{cui_fully_2011,Efthimiou2022}.

The design of \texttt{parallelproj} allows it to be integrated into other reconstruction platforms,
as illustrated here for STIR. 
However, for optimal performance, a re-design of the reconstruction platform might be required, as noted in section \ref{sec:STIR}. As shown in this paper, avoiding the overhead of copying data between CPU and GPU memory can have substantial impact. 
In C++, this could be avoided by using \texttt{CUDA} managed pointers, for instance via the CuVec library\footnote{\url{https://amypad.github.io/CuVec}}. 
However, best performance requires implementing most operations such as numerical algebra and filtering directly in \texttt{CUDA}, as illustrated in this paper.

\added{It is noteworthy that the current implementation of \texttt{parallelproj}'s
      Joseph projectors using arrays of LOR start and end coordinates is optimized towards (arbitrary) PET geometries.
      To calculate projections for reconstructing CT data acquired with a single moving source and a moving
      detector panel, more efficient implementations exploiting the known geometry between source and detector panel
      are possible.}

\added{Last but not least, it is worth highlighting that the python interface of \texttt{parallelproj} is compatible with the Python array API standard,
      enabling efficient projections and back projections of various compatible array classes (e.g. \texttt{numpy} CPU arrays,
      \texttt{cupy} GPU arrays, \texttt{pytorch} CPU and GPU tensors).
      This allows for a seamless integration of \texttt{parallelproj} into deep learning frameworks such as pytorch \cite{pytorch_2019}
      for the development of neural networks including forward and back projection layers such as unrolled variational networks \cite{Mehranian2021,Adler2018}.}

\section{Conclusion}

\texttt{parallelproj} is an open-source, and easy accessible research framework for efficient 
calculation of non-TOF and TOF projections in sinogram or listmode on multiple CPUs or 
state-of-the-art \texttt{CUDA} GPUs.
Conventional and advanced research reconstructions (including deep learning)
can be substantially accelerated by using the hybrid and pure GPU compute modes of this framework.

\section*{Acknowledgements}

The integration of \texttt{parallelproj} into STIR and SIRF was funded by the UKRI EPSRC grant to the 
Collaborative Computational Platform in Synergistic Reconstruction for Biomedical Imaging (EP/T026693/1).
The authors would like to thank Dr. Koen Michielsen for discussions on the Python array API standard.

  \printbibliography
\end{multicols}

\end{document}